\documentclass[12pt,a4paper]{article}

\usepackage{ifthen} 
\newboolean{pdflatex}
\setboolean{pdflatex}{true} 

\newboolean{articletitles}
\setboolean{articletitles}{true} 

\newboolean{uprightparticles}
\setboolean{uprightparticles}{false} 

\newboolean{inbibliography}
\setboolean{inbibliography}{false} 

\usepackage{microtype}
\usepackage{lineno}  
\usepackage{xspace} 
\usepackage{caption} 

\usepackage{graphicx}  
\usepackage{color}
\usepackage{colortbl}
\graphicspath{{./figs/}} 

\usepackage{amsmath} 
\usepackage{amssymb}
\usepackage{amsfonts}
\usepackage{upgreek} 

\newcommand*\patchAmsMathEnvironmentForLineno[1]{%
\expandafter\let\csname old#1\expandafter\endcsname\csname #1\endcsname
\expandafter\let\csname oldend#1\expandafter\endcsname\csname
end#1\endcsname
 \renewenvironment{#1}%
   {\linenomath\csname old#1\endcsname}%
   {\csname oldend#1\endcsname\endlinenomath}%
}
\newcommand*\patchBothAmsMathEnvironmentsForLineno[1]{%
  \patchAmsMathEnvironmentForLineno{#1}%
  \patchAmsMathEnvironmentForLineno{#1*}%
}
\AtBeginDocument{%
\patchBothAmsMathEnvironmentsForLineno{equation}%
\patchBothAmsMathEnvironmentsForLineno{align}%
\patchBothAmsMathEnvironmentsForLineno{flalign}%
\patchBothAmsMathEnvironmentsForLineno{alignat}%
\patchBothAmsMathEnvironmentsForLineno{gather}%
\patchBothAmsMathEnvironmentsForLineno{multline}%
\patchBothAmsMathEnvironmentsForLineno{eqnarray}%
}

\usepackage{hyperref}    
\usepackage[all]{hypcap} 

\def\lhcb {\mbox{LHCb}\xspace}

\def\MagUp {\mbox{\em Mag\kern -0.05em Up}\xspace}

\ifthenelse{\boolean{uprightparticles}}%
{

 \def\Ppi         {\ensuremath{\uppi}\xspace}

 \def\PDelta      {\ensuremath{\Delta}\xspace}                 
 \def\PXi      {\ensuremath{\Xi}\xspace}                 
 \def\PLambda      {\ensuremath{\Lambda}\xspace}                 
 \def\PSigma      {\ensuremath{\Sigma}\xspace}                 
 \def\POmega      {\ensuremath{\Omega}\xspace}                 
 \def\PUpsilon      {\ensuremath{\Upsilon}\xspace}                 
 
 \def\PB      {\ensuremath{\mathrm{B}}\xspace}                 
                  
 \def\PD      {\ensuremath{\mathrm{D}}\xspace}

 \def\PK      {\ensuremath{\mathrm{K}}\xspace}

 \def\Pi      {\ensuremath{\mathrm{i}}\xspace}

 \def\Ps      {\ensuremath{\mathrm{s}}\xspace}

}
{

 \def\Ppi         {\ensuremath{\pi}\xspace}

 \mathchardef\PDelta="7101
 \mathchardef\PXi="7104
 \mathchardef\PLambda="7103
 \mathchardef\PSigma="7106
 \mathchardef\POmega="710A
 \mathchardef\PUpsilon="7107
                  
 \def\PB      {\ensuremath{B}\xspace}                 
                  
 \def\PD      {\ensuremath{D}\xspace}

 \def\PK      {\ensuremath{K}\xspace}

 \def\Pi      {\ensuremath{i}\xspace}

 \def\Ps      {\ensuremath{s}\xspace}

}

\makeatletter
\ifcase \@ptsize \relax
  \newcommand{\miniscule}{\@setfontsize\miniscule{4}{5}}
\or
  \newcommand{\miniscule}{\@setfontsize\miniscule{5}{6}}
\or
  \newcommand{\miniscule}{\@setfontsize\miniscule{5}{6}}
\fi
\makeatother

\DeclareRobustCommand{\optbar}[1]{\shortstack{{\miniscule (\rule[.5ex]{1.25em}{.18mm})}
  \\ [-.7ex] $#1$}}

\def\squark    {{\ensuremath{\Ps}}\xspace}

\def\pion   {{\ensuremath{\Ppi}}\xspace}

\def\pip    {{\ensuremath{\pion^+}}\xspace}

\def\kaon    {{\ensuremath{\PK}}\xspace}
  \def\Kbar    {{\kern 0.2em\overline{\kern -0.2em \PK}{}}\xspace}

\def\KorKbar    {\kern 0.18em\optbar{\kern -0.18em K}{}\xspace}

\def\Kp      {{\ensuremath{\kaon^+}}\xspace}
\def\Km      {{\ensuremath{\kaon^-}}\xspace}

\def\KS      {{\ensuremath{\kaon^0_{\rm\scriptscriptstyle S}}}\xspace}

  \def\Dbar    {{\kern 0.2em\overline{\kern -0.2em \PD}{}}\xspace}
\def\D       {{\ensuremath{\PD}}\xspace}

\def\DorDbar    {\kern 0.18em\optbar{\kern -0.18em D}{}\xspace}
\def\Dz      {{\ensuremath{\D^0}}\xspace}

\def\Dp      {{\ensuremath{\D^+}}\xspace}

\def\Dstarp  {{\ensuremath{\D^{*+}}}\xspace}

\def\Ds      {{\ensuremath{\D^+_\squark}}\xspace}

\def\Bbar    {{\ensuremath{\kern 0.18em\overline{\kern -0.18em \PB}{}}}\xspace}

\def\BorBbar    {\kern 0.18em\optbar{\kern -0.18em B}{}\xspace}

  \def\Y#1S{\ensuremath{\PUpsilon{(#1S)}}\xspace}

\def\Lbar        {{\ensuremath{\kern 0.1em\overline{\kern -0.1em\PLambda}}}\xspace}
\def\LorLbar    {\kern 0.18em\optbar{\kern -0.18em \PLambda}{}\xspace}

\def\to                 {\ensuremath{\rightarrow}\xspace}

\def\CP                {{\ensuremath{C\!P}}\xspace}

\def\AT#1     {\ensuremath{A_{\mathrm{T}}^{#1}}\xspace}           

\def\C#1      {\ensuremath{\mathcal{C}_{#1}}\xspace}                       
\def\Cp#1     {\ensuremath{\mathcal{C}_{#1}^{'}}\xspace}                    
\def\Ceff#1   {\ensuremath{\mathcal{C}_{#1}^{\mathrm{(eff)}}}\xspace}        
\def\Cpeff#1  {\ensuremath{\mathcal{C}_{#1}^{'\mathrm{(eff)}}}\xspace}       
\def\Ope#1    {\ensuremath{\mathcal{O}_{#1}}\xspace}                       
\def\Opep#1   {\ensuremath{\mathcal{O}_{#1}^{'}}\xspace}                    

\newcommand{\tev}{\ifthenelse{\boolean{inbibliography}}{\ensuremath{~T\kern -0.05em eV}\xspace}{\ensuremath{\mathrm{\,Te\kern -0.1em V}}}\xspace}
\newcommand{\gev}{\ensuremath{\mathrm{\,Ge\kern -0.1em V}}\xspace}
\newcommand{\mev}{\ensuremath{\mathrm{\,Me\kern -0.1em V}}\xspace}
\newcommand{\kev}{\ensuremath{\mathrm{\,ke\kern -0.1em V}}\xspace}
\newcommand{\ev}{\ensuremath{\mathrm{\,e\kern -0.1em V}}\xspace}
\newcommand{\gevc}{\ensuremath{{\mathrm{\,Ge\kern -0.1em V\!/}c}}\xspace}
\newcommand{\mevc}{\ensuremath{{\mathrm{\,Me\kern -0.1em V\!/}c}}\xspace}
\newcommand{\gevcc}{\ensuremath{{\mathrm{\,Ge\kern -0.1em V\!/}c^2}}\xspace}
\newcommand{\gevgevcccc}{\ensuremath{{\mathrm{\,Ge\kern -0.1em V^2\!/}c^4}}\xspace}
\newcommand{\mevcc}{\ensuremath{{\mathrm{\,Me\kern -0.1em V\!/}c^2}}\xspace}

\def\invpb {\ensuremath{\mbox{\,pb}^{-1}}\xspace}

\def\khz  {\ensuremath{{\rm \,kHz}}\xspace}

\newcommand{\chisq}{\ensuremath{\chi^2}\xspace}

\def\gsim{{~\raise.15em\hbox{$>$}\kern-.85em
          \lower.35em\hbox{$\sim$}~}\xspace}
\def\lsim{{~\raise.15em\hbox{$<$}\kern-.85em
          \lower.35em\hbox{$\sim$}~}\xspace}
\def\mysim{\ensuremath\sim\kern-0.3em\xspace}

\def\tell1  {TELL1\xspace}
\def\ukl1   {UKL1\xspace}

\usepackage{cite} 
\usepackage{mciteplus}

\usepackage{longtable} 

\begin{document}

\renewcommand{\thefootnote}{\fnsymbol{footnote}}
\setcounter{footnote}{1}


\begin{titlepage}
\pagenumbering{roman}

\vspace*{-1.5cm}
\centerline{\large EUROPEAN ORGANIZATION FOR NUCLEAR RESEARCH (CERN)}
\vspace*{1.5cm}
\noindent
\begin{tabular*}{\linewidth}{lc@{\extracolsep{\fill}}r@{\extracolsep{0pt}}}
\ifthenelse{\boolean{pdflatex}}
{\vspace*{-2.7cm}\mbox{\!\!\!\includegraphics[width=.14\textwidth]{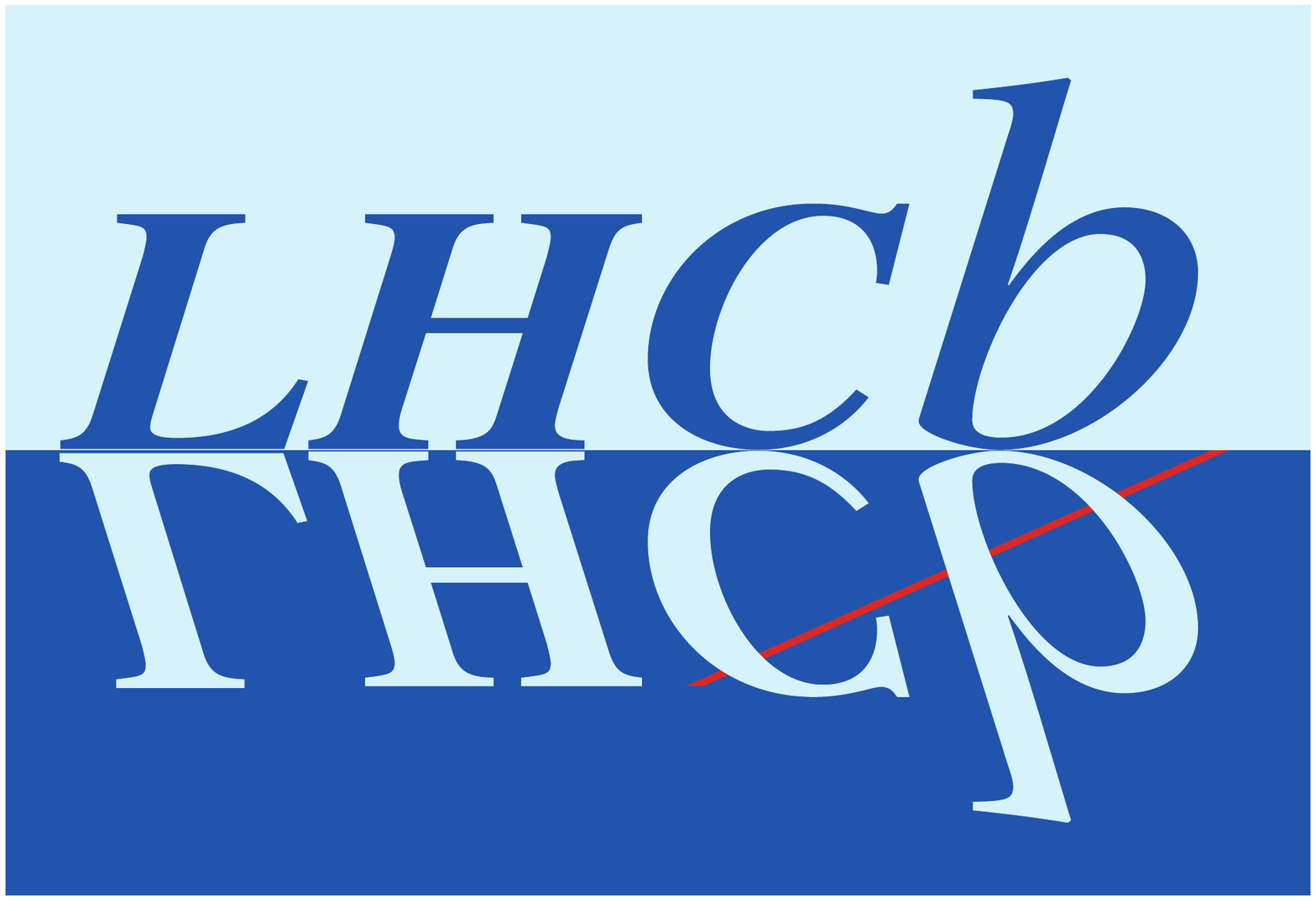}} & &}%
{\vspace*{-1.2cm}\mbox{\!\!\!\includegraphics[width=.12\textwidth]{lhcb-logo.eps}} & &}%
\\
 & & 19 April 2016 \\ 
 & & CERN-LHCb-DP-2016-001 \\
\end{tabular*}

\vspace*{1.0cm}

{\bf\boldmath\huge
\begin{center}
Tesla : an application for real-time data analysis in High Energy Physics
\end{center}
}

\vspace*{0.0cm}

\begin{center}
\small
R. Aaij$^{1}$, S. Amato$^{2}$, L. Anderlini$^{3}$, S. Benson$^{1}$, M. Cattaneo$^{1}$, M. Clemencic$^{1}$, B. Couturier$^{1}$,
M. Frank$^{1}$, V.V. Gligorov$^{4}$, T. Head$^{5}$, C. Jones$^{6}$, I. Komarov$^{5}$, O. Lupton$^{7}$, R. Matev$^{1}$, G. Raven$^{8}$, 
B. Sciascia$^{9}$, T. Skwarnicki$^{10}$, P. Spradlin$^{11}$, S. Stahl$^{1}$, B. Storaci$^{12}$, 
M. Vesterinen$^{13}$.\\
\bigskip
{\footnotesize \it
$ ^{1}$European Organization for Nuclear Research (CERN), Geneva, Switzerland\\
$ ^{2}$Universidade Federal do Rio de Janeiro (UFRJ), Rio de Janeiro, Brazil\\
$ ^{3}$Sezione INFN di Firenze, Firenze, Italy\\
$ ^{4}$LPNHE, Universit\'{e} Pierre et Marie Curie, Universit\'{e} Paris Diderot, CNRS/IN2P3, Paris, France\\
$ ^{5}$Ecole Polytechnique F\'{e}d\'{e}rale de Lausanne (EPFL), Lausanne, Switzerland\\
$ ^{6}$Cavendish Laboratory, University of Cambridge, Cambridge, United Kingdom\\
$ ^{7}$Department of Physics, University of Oxford, Oxford, United Kingdom\\
$ ^{8}$Nikhef National Institute for Subatomic Physics and VU University Amsterdam, Amsterdam, The Netherlands\\
$ ^{9}$Laboratori Nazionali dell'INFN di Frascati, Frascati, Italy\\
$ ^{10}$Syracuse University, Syracuse, NY, United States\\
$ ^{11}$School of Physics and Astronomy, University of Glasgow, Glasgow, United Kingdom\\
$ ^{12}$Physik-Institut, Universit\"{a}t Z\"{u}rich, Z\"{u}rich, Switzerland\\
$ ^{13}$Physikalisches Institut, Ruprecht-Karls-Universit\"{a}t Heidelberg, Heidelberg, Germany\\
}
\end{center}

\vspace{\fill}

\begin{abstract}
  \noindent
  Upgrades to the \lhcb computing infrastructure in the first long shutdown of the LHC have allowed
  for high quality decay information to be calculated by the software trigger
  making a separate offline event reconstruction unnecessary.
  Furthermore, the storage space of the triggered candidate is an order of magnitude smaller than the entire raw event 
  that would otherwise need to be persisted. Tesla, following the \lhcb renowned physicist naming convention,
  is an application designed to process the information calculated by the trigger, with the resulting
  output used to directly perform physics measurements.
\end{abstract}

\vspace{\fill}

\begin{center}
  Submitted to Journal of Computational Physics
\end{center}

\vspace{\fill}

{\footnotesize 
\centerline{\copyright~CERN on behalf of the \lhcb collaboration, licence \href{http://creativecommons.org/licenses/by/4.0/}{CC-BY-4.0}.}}
\vspace*{2mm}

\end{titlepage}


\newpage
\setcounter{page}{2}
\mbox{~}
%
%
%
%

\cleardoublepage


\renewcommand{\thefootnote}{\arabic{footnote}}
\setcounter{footnote}{0}



\pagestyle{plain} 
\setcounter{page}{1}
\pagenumbering{arabic}

%

\section{Introduction}
\label{sec:intro}

The \lhcb experiment, one of the four main detectors situated on the Large Hadron Collider in CERN, Geneva, specialises
in precision measurements of beauty and charm hadrons decays.
Large backgrounds are present at hadron colliders. 
At the nominal LHCb luminosity of $4\times10^{32}\, {\rm cm}^{-2}{\rm s}^{-1}$ during 2012 data taking at 8\tev, 
around 30\,k beauty ($b$) and 600\,k charm ($c$) hadron pairs pass through the detector each second. 
Each recorded collision is defined as an event that can possibly contain a decay of interest.
The most interesting $b$-hadron decays typically occur with decay probabilities of less than $10^{-5}$, whereas
a large fraction of $c$-hadron decays are retained for further study.
The efficient selection of beauty and charm decays from the $\mysim30$\,M proton-proton collisions per second is a significant Big Data challenge.

An innovative feature of the 
LHCb experiment is its approach to Big Data in the form of the High Level 
Trigger (HLT)~\cite{LHCb-DP-2012-004}. This is a software application designed to reduce the event
rate from 1\,M to $\mysim10$\,k events per second and is executed on an Event Filter 
Farm (EFF). The EFF is a computing cluster consisting of 1800 server nodes, with a combined storage space of 
5.2~PB. This can accommodate up to two weeks of LHCb data taking~\cite{LHCb-DP-2014-002} in nominal conditions. 
The HLT application reconstructs the particle trajectories of the event in real time, where real time is defined as the interval
between the collision in the detector and the moment the data are sent to permanent storage. The event reconstruction
in the EFF is denoted as the \emph{online reconstruction}.

In the LHCb data processing model of LHC Run-I (2010-2012), all events accepted by the 
HLT were sent to permanent offline storage containing all raw information from the detector.
An additional event reconstruction performed on the LHC Computing Grid~\cite{Stagni:2012nz}, denoted the \emph{offline reconstruction},
recreates particles in the event from the raw data using an improved detector calibration.

The upgrade of the computing infrastructure during the first long shutdown of the LHC (2013-2014), 
combined with efficient use of the EFF storage, provides resources for an online reconstruction in LHC Run-II (2015-2018) with a similar 
quality to that of the offline reconstruction.
This is achieved through real-time automated calculation of the final calibrations of the sub-detectors. 

With offline-quality information available at the HLT level,
it is possible to perform physics analyses with the information calculated by the HLT event reconstruction.
In the {\em Turbo stream}, a compact event record is written directly 
from the trigger and is prepared for physics analysis by the Tesla application.
This bypasses the offline reconstruction.
Reaching the ultimate precision of the \lhcb experiment already in real time as the data arrive
has the power to transform the experimental approach to 
processing large quantities of data. 

The data acquisition framework is described in Section~\ref{sec:trigger}.
An overview of the upgrades to the trigger and calibration
framework in Run-II is provided in Section~\ref{sec:run2}.
The implementation of the Turbo stream including that of the Tesla application is described in Section~\ref{sec:implementation},
followed by the future prospects of the data model in Section~\ref{sec:future}.

\section{The LHCb detector, data acquisition, and trigger strategy}
\label{sec:trigger}

The \lhcb detector is a forward arm spectrometer
designed to measure the properties of the decays of $b$-hadrons with
high precision~\cite{Alves:2008zz}. Such decays are predominantly produced
at small angles with respect to the proton beam axis~\cite{LHCb-PAPER-2010-002}. 
This precision is obtained with an advanced tracking system consisting
of a silicon vertex detector surrounding the interaction region (VELO), a silicon strip
detector located upstream of the dipole magnet (TT), and three tracking stations downstream of the
magnet, which consist of silicon strip detectors in the high intensity region close to the
beamline (IT) and a straw-tube tracker in the regions further from the beamline (OT).
Neutral particles are identified with a calorimeter system consisting of a scintillating
pad detector (SPD), an electromagnetic calorimeter preceded by a pre-shower detector (ECAL, PS),
and a hadronic calorimeter (HCAL). Particle identification is provided by combining information from the ring-imaging Cherenkov
detectors (RICH1 and RICH2), the wire chambers used to detect muons, and the calorimeter system.

\subsection{Data readout and hardware trigger}

Most interesting physics processes occur at event rates of 
less than 10~Hz in Run-I conditions.
This can be compared to the 30~MHz at which bunches of protons are brought to collision.
Reducing the output rate through the use of a trigger system is
essential to reject uninteresting collisions, thereby using computing resources more efficiently.
The detector front-end (FE) electronic boards connect to a set of common readout boards (RB) that limit the event output rate to 1.1\,MHz.
At LHCb a three-level trigger system is used consisting of a hardware level followed by two software levels. 
The level-0 hardware trigger (L0) reduces the 
input collision rate of $\mysim30$\,MHz to the maximum output rate allowed by the front-end electronics.
The L0 decision is based on algorithms running on field-programmable gate arrays (FPGAs), 
which achieve the necessary reduction in rate within a fixed latency of $4\,\mu {\rm s}$. 
Information from the ECAL, HCAL, and muon stations
is used in FPGA calculations in separate L0 algorithms. Decisions from these different hardware
triggers are combined and passed to the readout supervisor (RS). 
The readout boards perform zero-suppression and interface the 
custom electronics to the readout network via Gigabit Ethernet links.
The RS decides where in the EFF to send the event based on the state
of the buffers and the available resources in the EFF. The EFF node address information is sent to the RB via optical
links that also keeps them synchronised.
Event raw data are distributed among the individual servers of the EFF using a simple credit-request scheme.

\subsection{High Level Trigger}
\label{sec:HLT}

The LHCb high level trigger (HLT) is a software application, executed on the EFF, that is implemented 
in the same Gaudi framework~\cite{Barrand:2001ny} as the software used for the offline reconstruction. 
This permits the incorporation of offline reconstruction software in the trigger,
provided that it is sufficiently optimised to be used in real time.
While no sub-detector upgrade took place in the first long shutdown of the LHC (2013-2014), the EFF was improved.
The EFF now consists of approximately 1800 nodes, with 1000 containing 2~TB of hard disk space each
and 800 nodes containing 4TB each, giving a total of $5.2$~PB. 
Each server node in the EFF contains 12-16 physical processor cores 
and 24-32 logical cores.

The first level of the software trigger (HLT1) reconstructs charged particle trajectories 
using information from the VELO and tracking stations.
If at least one track is found that satisfies strict quality and transverse momentum criteria, then
the event is passed to the second level of the software trigger (di-muon combinations may also trigger the first
software level). The output rate of HLT1 is $\mysim150$\khz.

The second level of the software trigger (HLT2) can use information from
all sub-detectors to decide whether or not to keep an event for analysis and permanent storage.
A full event reconstruction is performed such that HLT2 is then able to identify the most interesting events,
with a final output rate of $\mysim10$\khz.

\section{Run-II data taking}
\label{sec:run2}

During Run-I data taking, a buffer was created between the hardware trigger and
the first software trigger level, deferring 20\,\% of the events passing the hardware trigger and 
thereby utililising the EFF when the LHC was not providing proton collisions.
The replacement of this buffer with the one between the two software levels introduces
a complete split between an initial stage processing events directly from the L0 (HLT1) and an
asynchronous stage (HLT2), ensuring the EFF is used optimally.

\begin{figure}[t]
\begin{center}
\includegraphics[width=0.82\textwidth]{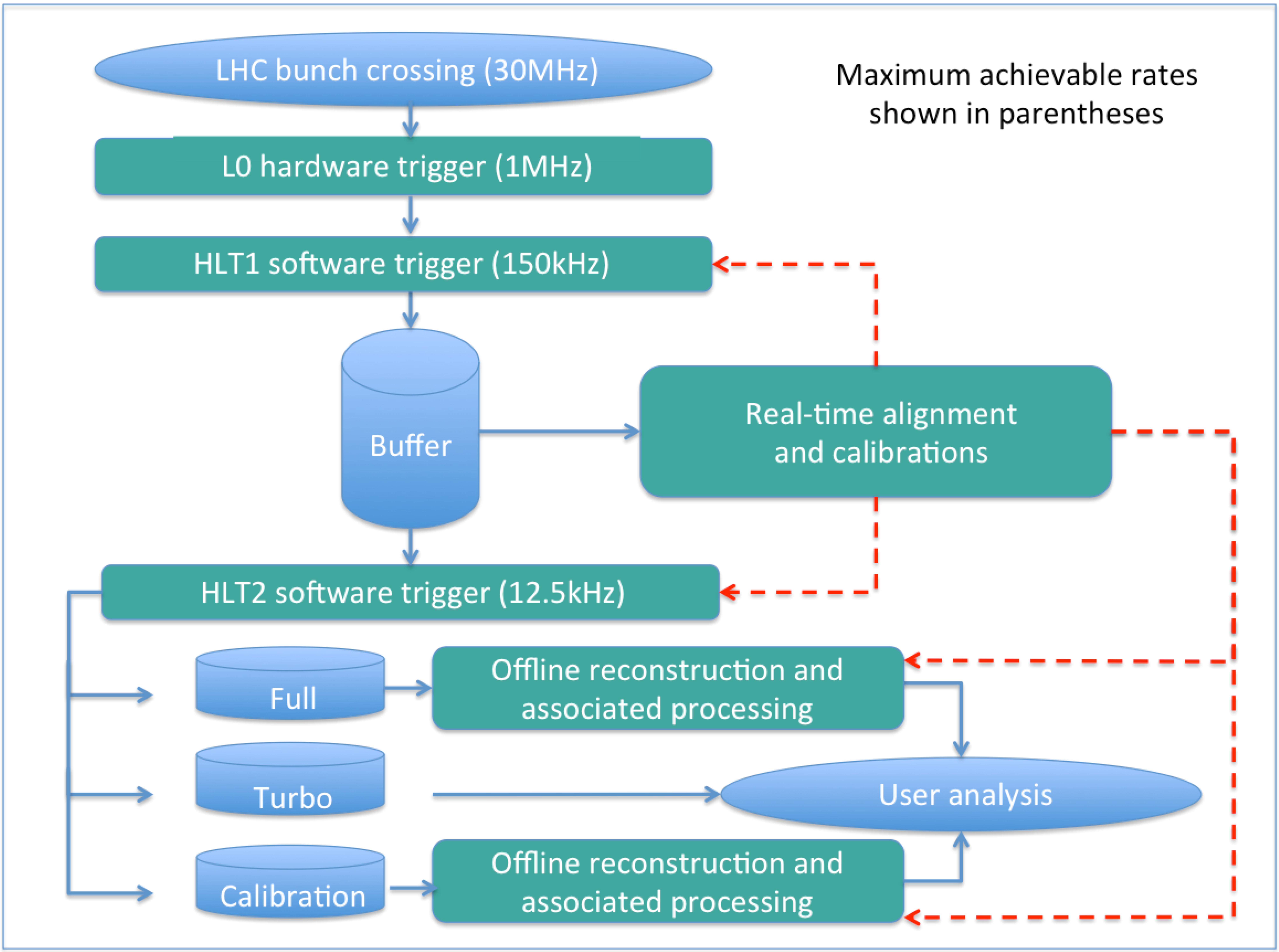}
\caption{Schematic diagram showing the overall data processing model in Run-II,
where the blue solid line represents data flow, and the red dashed line 
the propagation of calibrations.}
\label{fig:trigger2015}
\end{center}
\end{figure}

From 2015 onwards, the two HLT software levels run independently,
giving rise to the Run-II data flow depicted in Figure~\ref{fig:trigger2015}.
In order to process two independent trigger stages, a substantial modification to
the trigger software was required (a detailed description is provided in Ref.~\cite{Frank:2014ixa}).
This included the creation of an additional buffer between the two software levels.
The flexibility in trigger processing that is provided by this buffer system allows 
the execution of high-quality alignment and calibration between HLT1 and HLT2.

The alignment and calibrations are 
described in detail Section~\ref{sec:align}.
The different data streams created from the HLT2 selected events are 
described in Section~\ref{sec:streaming}. Among these is the Turbo stream in which information from the
online reconstruction is persisted, detailed in Section~\ref{sec:implementation}.

\subsection{Real-time alignment and calibration}
\label{sec:align}

As described in Section~\ref{sec:HLT}, increased computing resources in the EFF allow for automated
alignment and calibration tasks to supply high-quality information to the trigger software.
This removes the need for further reprocessing.

In order to align and calibrate the detector, dedicated samples from 
HLT1 are taken as input. These calibrations are implemented before HLT2 has processed the data, which is a prerequisite
for the successful functioning of the Turbo stream as it relies on information calculated by HLT2. The calibrations
are also used by HLT1 for subsequent processing. The alignment and calibration tasks are performed at regular intervals. These intervals
can be as frequent as each time a new LHC proton beam is formed or less frequent depending on the calibrations being calculated. 
The calibration tasks are performed in a few minutes using the nodes from the EFF. The resulting
alignment or calibration parameters are updated if they differ significantly from the values
used at a given time and are stored in a database as a function of time.

Full details of the real-time alignment procedure are provided in Ref.~\cite{Dujany:2015} and
are summarised in this section.
The major detector alignment and calibration tasks consist of the following.

\textit{Alignment of the VELO and tracking stations.} Misalignment of the VELO and tracking stations
has a direct impact on the momentum resolution of charged particles.
The alignment is achieved through minimisation of the residuals of a Kalman fit~\cite{Fruhwirth:1987fm}
to a set of well reconstructed tracks from HLT1.
The alignment of the VELO is performed first followed by the corresponding
alignment of the tracking stations. This order is chosen due to the nature of the VELO,
which centres itself around the collision point\footnote{Typically beams are recreated at approximately 8 hour intervals.}.

\textit{RICH mirror alignment.} Misalignment of the RICH detectors causes the circular
rings on the detection plane to become distorted. Therefore
the distance from the projected position of the track to the Cherenkov ring varies as a
function of the azimuthal angle. This distortion requires an individual correction for each mirror. The
correction is calculated using a set of well reconstructed tracks from HLT1.

\textit{Global time alignment of the OT.} The outer tracker uses a drift (gas) tube design~\cite{Arink:2013twa}.
The measurement of the drift time in the straw tubes
is susceptible to differences between the true collision time and the LHCb clock. Such a difference causes the
measured drift time to be different from the estimated time arising from the distance of the wire to the track.
The residuals of a sample of well reconstructed tracks are used to provide
a global drift time offset for the tubes.

\textit{RICH radiator refractive index calibration.} Gas radiators in the RICH detectors
are the source of Cherenkov radiation. The refractive index of the gas varies as a function of 
temperature and pressure, which change over time. From a set of well reconstructed tracks originating from a particle of known mass
the expected Cherenkov angle can be calculated using accurate momentum measurements provided by the tracking stations.
The distribution of expected versus measured Cherenkov angles provides the basis for the refractive index calibration
of the radiator.

The complete calibration of the full detector is a complex enterprise.
The achievement of automating and providing accurate calibrations
within a few minutes is a substantial achievement without which analysis-quality reconstruction in HLT2 would be impossible.

\subsection{Streaming and data flow}
\label{sec:streaming}

Data streams at LHCb are controlled through the assignment of {\em routing bits} to each event by the trigger
software.
Routing bits are set according to the trigger lines that select the event.
A filter is applied based on the routing bit that allows for
different streams to be sent to different places in permanent storage, as depicted in Figure~\ref{fig:trigger2015}.
These data are processed
with the DIRAC software framework~\cite{Stagni:2012nz}, which is used to manage all LHCb data processing on the Grid.
The physics data streams in Run-II are the Full stream, the Turbo stream,
and the Calibration stream. Events are allowed to be present
in more than one stream.

Events that are sent to the Full stream have luminosity information calculated, which is then stored in the data
file as a file summary record (FSR). They undergo a further offline reconstruction using the sub-detector
raw data banks, which contain the detector information. Subsequent analysis
selections are applied, identifying decay channels of interest. After this processing is completed, files
are merged to improve network utilisation in the LHC grid
computing infrastructure, and the data are then available for physics analyses.

The events directed to the Turbo stream consist
of the flagged physics lines and those triggered for luminosity accounting.
The Turbo stream does not require further event reconstruction, since datasets are ready for user analysis directly
after the generation of luminosity file summary records and the restoration of the trigger objects.
As for the Full stream the resulting files are merged.
This is illustrated in Figure~\ref{fig:turbo2015}. 

\subsubsection*{Calibration}

Calibration candidates provide pure data samples of pions, kaons, protons, 
muons and electrons that can be used to determine efficiencies in a data-driven method.
Most LHCb analyses apply selection requirements for variables that show some
disagreement between real data and simulated data, for example Particle Identification.
The determination of efficiencies using data is therefore preferred where possible.

Exclusive selections applied in the trigger allow candidates of highly abundant 
unstable particles, such as $\Dstarp\to\Dz\pip$ to be directed to a dedicated stream.
Events sent to the Calibration stream contain the stored 
trigger candidates in addition to the raw sub-detector data. 
Thus both the offline reconstruction of the Full stream and the 
trigger restoration of the Turbo stream can be applied to Calibration stream events.
The workflow of the Calibration stream is depicted in Figure~\ref{fig:turbo2015}.
In this way the same particle decay candidates can be used to provide data driven corrections for 
both the online and offline event reconstructions.

\section{Implementation of the Turbo stream}
\label{sec:implementation}

\begin{figure}
\begin{center}
\includegraphics[width=0.75\textwidth]{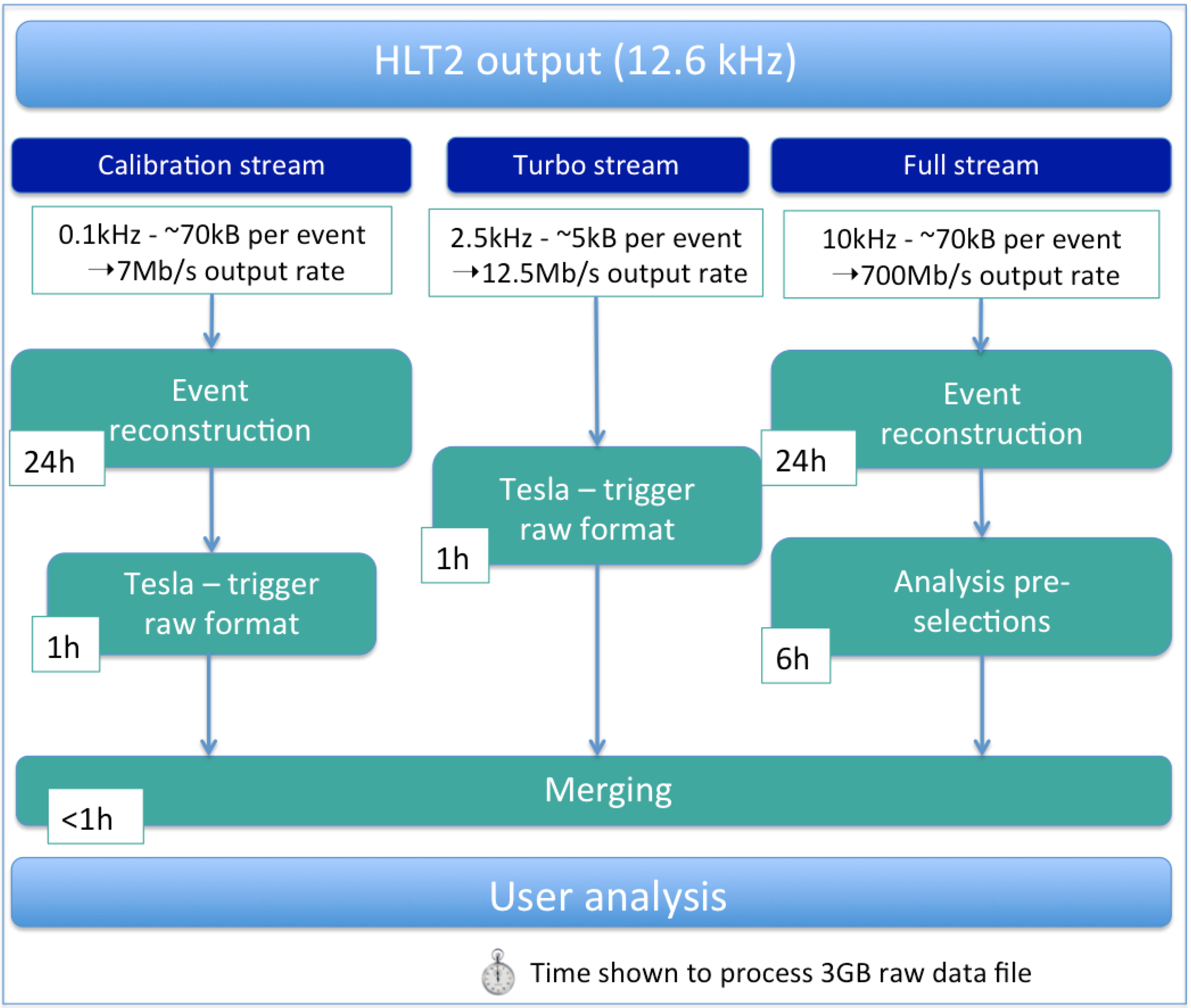}
\caption{Turbo data processing versus the traditional approach,
as described in Section~\ref{sec:implementation}. The time taken for each step in hours is
provided for a 3\,GB raw data file. In addition, a calibration stream
separates events for further processing to calculate data-driven efficiencies for both the Full and
Turbo streams.}
\label{fig:turbo2015}
\end{center}
\end{figure}

The concept of the Turbo stream is to provide a framework by which
physics analyses can be performed using the online reconstruction directly.
The schematic data flow of the Turbo stream compared to the traditional 
data flow (represented by the Full stream) is depicted in Figure~\ref{fig:turbo2015}.

In the traditional data flow, raw event data undergoes a
complete reconstruction taking 24 hours for 3\,GB of input data on a typical batch node.
This additional reconstruction was designed for a data processing model in which
final calibrations performed at the end of a data taking year were a significant
improvement compared to the calibrations initially available. This is no longer needed since
high quality calibrations are produced in real time.
After the offline reconstruction, selection criteria based on typical $b$-hadron and $c$-hadron topologies are applied that
identify the decay candidates for user analysis, taking an additional 6 hours for the
3\,GB of raw data. After a final merging step of less than one hour, the datasets
are ready for user analysis. The approach taken by the Turbo stream is to save the particle
candidates reconstructed in the HLT (equivalent to those produced after the 
selection stage in the traditional approach) inside the raw event. The time taken
for the Tesla application to format the data in preparation for user analysis is
approximately 1 hour.

A clear advantage of the Turbo stream is that the event size is an order
of magnitude smaller than that of the Full stream as all sub-detector information may be
discarded. 
For standard Run-II conditions $\mysim20\,\%$ of the HLT2 selected events will be sent to the Turbo stream 
at a cost of less than 2\,\% of the output bandwidth.

In order to perform physics analyses with the online reconstruction, 
decay candidates must appear in the same format as expected by the output of the  traditional
processing, such that the existing analysis infrastructure can be used. This is
the purpose of the Tesla application. The high-level functions of the
Tesla application are detailed in Section~\ref{sec:tesla}. The low-level
design is described in Section~\ref{sec:storing}.

\subsection{The Tesla application}
\label{sec:tesla}

For events in the Turbo stream, the reconstructed decay candidates are stored in the raw data in the format
usually reserved for detector level information.
The Tesla application is subsequently used to perform multiple tasks on events sent to the Turbo stream.

The purpose of Tesla is to ensure that the resulting output
is in a format that is ready for analysis. 
This means that the Tesla application must do the following:
\begin{itemize}
\item Compute the information that is necessary for the luminosity determination
and store this in the output file, as described in Section~\ref{sec:streaming}.
\item Place the HLT candidate decays in the output file in such a way that existing analysis tools function correctly with minimal modifications.
\item Ensure that additional information calculated in the online reconstruction is accessible to standard analysis tools, for example
event-level sub-detector occupancies and information calculated using the whole reconstructed collision.
\item Discard the raw data corresponding to the sub-detectors, leaving only the information on the requested HLT candidates, trigger decisions and headers required for subsequent analysis. It should be
noted that this requirement is only relevant for 2015 commissioning and will take place
inside the trigger itself from 2016 onwards.
\end{itemize}
The Tesla application must capable of processing simulated data sets.  
Thus it must be able to associate reconstructed tracks and calorimeter clusters to the 
simulated particles that produced these signatures in the simulated detector.
There should be also an option to protect the detector raw data, such that the offline
reconstruction in the traditional data flow can coexist in simulated events alongside
the physics analysis objects from the online reconstruction.
It should be noted that the objects created by the Tesla application are stored in
different addresses in the output file than those
used by the traditional processing so that there can be no interference.
This is a requirement for the calibration stream as described in Section~\ref{sec:streaming}.

\subsubsection*{Monitoring and validation}

In addition to providing analysts with data for physics measurements, the Tesla application
allows for a new method with which to monitor the stability of the detector and the reconstruction.
Monitoring is traditionally performed at LHCb by studying abundant decays such as $\Dz\to\Km\pip$
using a separate implementation of the offline reconstruction. The data used is that of a single run, which is 1 hour in duration at most.
The creation of histograms directly from the Tesla application permits the simultaneous analysis of several runs.
\begin{figure}[t]
\begin{center}
\includegraphics[width=1.0\textwidth]{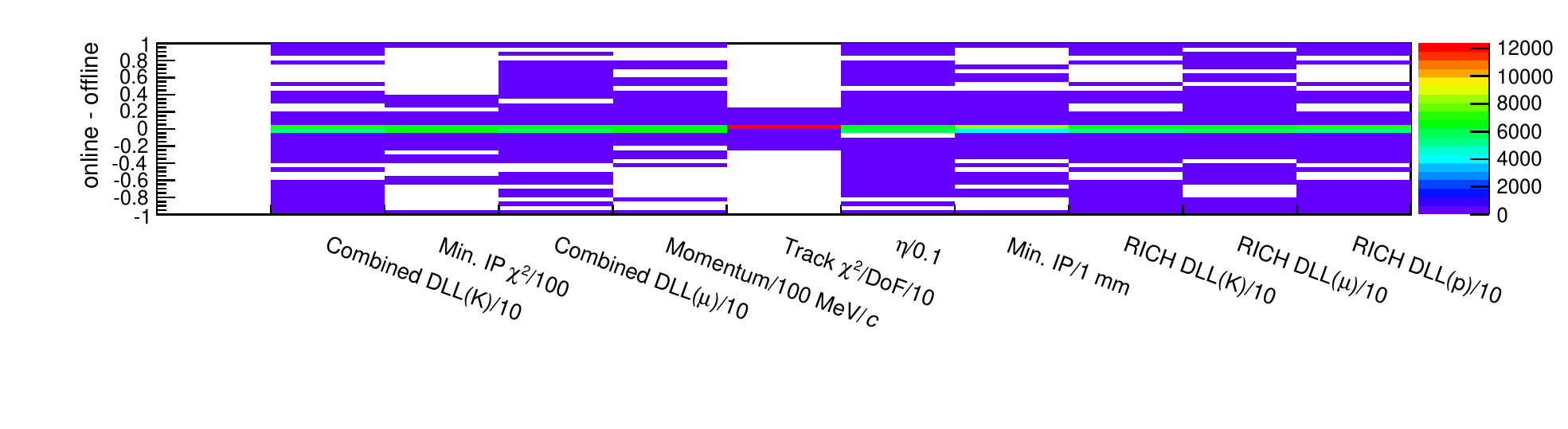}
\caption{
  Comparison between variables calculated by the online and offline event reconstructions
  for the pion in $\Dz\to\Km\pip$ decays of a single hour of LHCb data taking, where the colour represents
    the number of pion candidates. The variables are defined in the text.
}
\label{fig:comp}
\end{center}
\end{figure}
The additional monitoring allows for variables calculated in the online and the offline
event reconstructions to be compared to ensure consistency. 
This also serves the purpose of validating the use of the real-time alignment inside the offline processing.
An example comparison is given in Figure~\ref{fig:comp} of the reconstruction variables calculated for the pion in
$\Dz\to\Km\pip$ decays. The variables shown consist of the following. The difference in log-likelihood between a given particle type and 
that of a pion both combining all sub-detector information (Combined DLL) and restricting to the RICH sub-detector information (RICH DLL), 
the minimum difference in \chisq of the primary vertex between the fits with and without
the pion candidate (Min. IP \chisq), the track fit quality (Track \chisq/DoF), the reconstructed pseudorapidity ($\eta$) and momentum, together with the minimum impact parameter (Min. IP) 
with respect to the primary vertices.
Comparing the online and offline calculated variables of two decay candidates requires that they are associated
to the same detector hits thereby ensuring the same candidate is found.

\subsection{Persistence of trigger objects}
\label{sec:storing}

Measurements of \CP-violating asymmetries along with searches for new states and rare decays
are made using datasets that are produced from physics object classes, written in C++.
These are stored
in the hashed containers of the data summary tape (DST) format. 
In the Full stream, the size of the raw event is on average 70\,kB. 
When the sub-detector banks are removed from the raw event,
and replaced with the specific trigger candidates firing the Turbo stream triggers, the event size is decreased to around 5\,kB.

\begin{figure}[t]
\begin{center}
\includegraphics[width=0.6\textwidth]{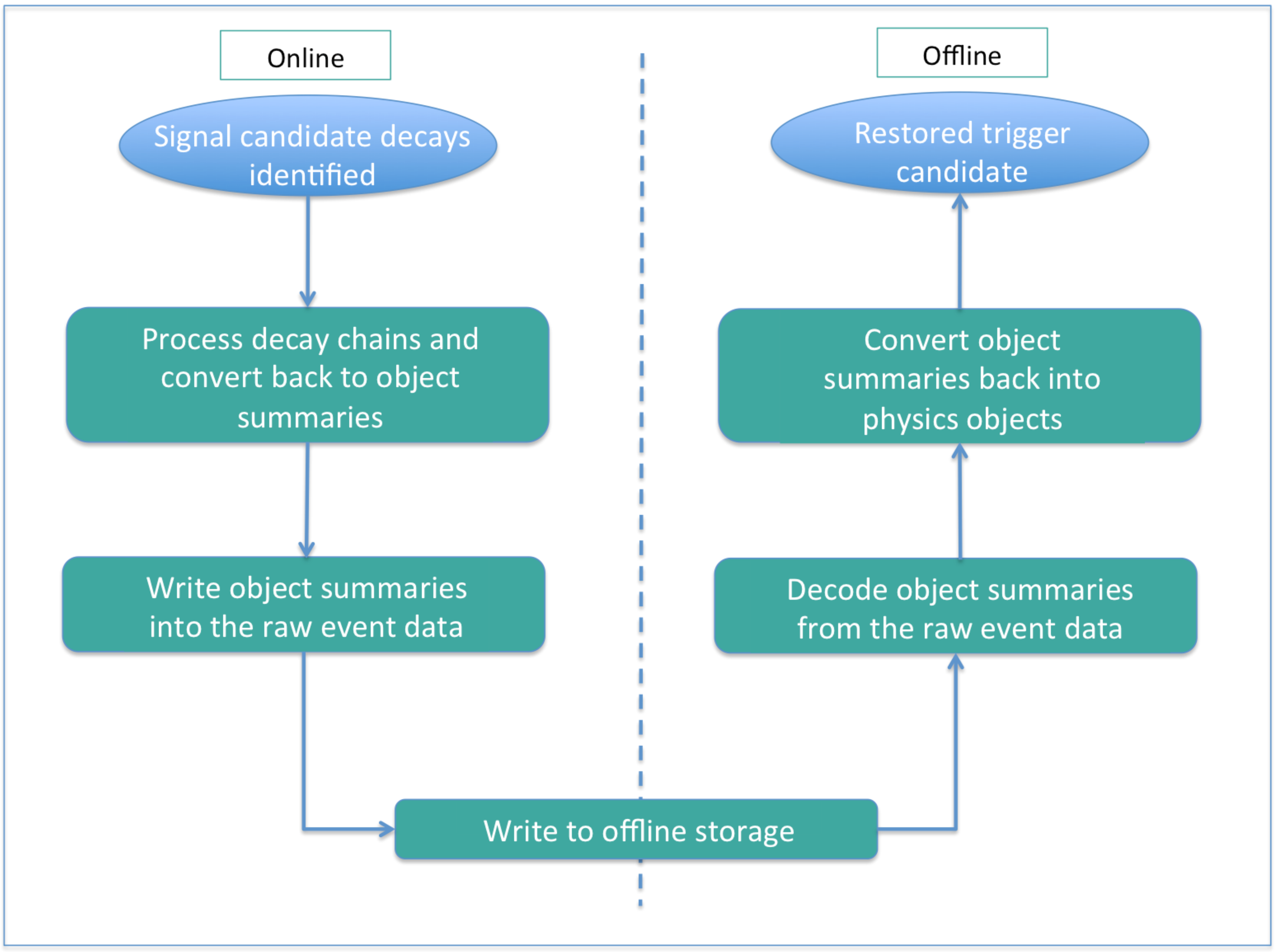}
\caption{Steps required to save and restore trigger objects.}
\label{fig:reports}
\end{center}
\end{figure}
During Run I data taking (2009-2012), some limited information from physics objects made inside the trigger
was already placed inside the raw event. 
This allows the possibility to determine trigger efficiencies directly from the data
using the methods described in Ref.~\cite{LHCb-DP-2012-004}.
The so-called selection reports allow for a C++ class, known as an HLT object summary, to save the 
members of any C++ physics object class in a key-value pair and point to other object summary instances. 
After the physics objects have been converted into object summaries, 
a dedicated algorithm writes them into the dedicated sub-bank
of the raw data bank associated to the HLT.
This is depicted in Figure~\ref{fig:reports}.

\begin{figure}[t]
\begin{center}
\includegraphics[width=0.82\textwidth]{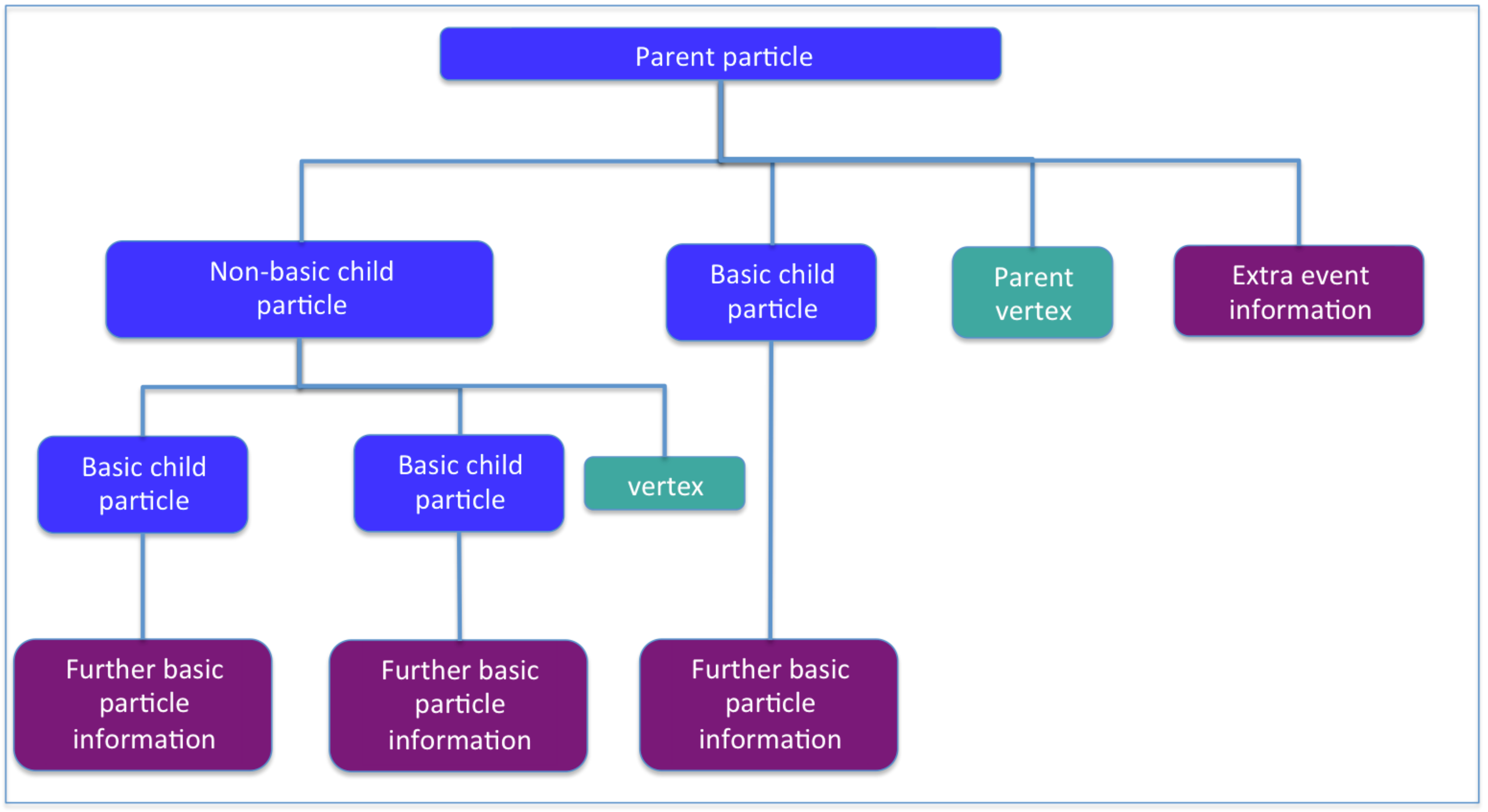}
\caption{Structure of stored objects. Each top-level object represents a
parent particle, which then points to child particles along with associated
information. Note that the example shown is for a two-level particle decay.
The infrastructure allows for as many levels as needed.}
\label{fig:pattern}
\end{center}
\end{figure}

A complete physics measurement
requires much more information than
was saved in the selection reports in Run I.
This means that many more classes must be placed into
the raw event and more information must be saved about each individual class. 
In order to save entire decay chains, a novel pattern is required
to describe the topology. 
This, combined with the information inside the summaries,
allows for all the required information of the decay to be saved. 
The pattern of the reports used for saving the decay topology is shown in Figure~\ref{fig:pattern}.

One of the main advantages of the method is its flexibility as only relevant information
contained in the object classes is persisted. Additional members added at a later date
can also be stored.
In order to preserve backwards compatibility, the version of the raw bank
is written to its header.
The version indicates to the dedicated converter tool exactly which piece of
information about the class is present for a given location in the raw data stream.
It is important that the restored data match the original data.
In order to achieve this, 
the same dedicated converter tool is used both to read and to write the object summary. For a given analysis class,
the position in the raw data stream is dedicated to a specific quantity. Individual
static maps for each analysis class, such as particles and tracks, are used by the
converter tool to keep track of exactly where each piece of information is stored.
The raw bank version is included on an event-by-event basis, which means that multiple versions can be 
processed simultaneously, avoiding the issue of backwards compatibility.

\subsection{Additional analysis requirements}

\begin{figure}[t]
\begin{center}
\includegraphics[width=0.45\textwidth]{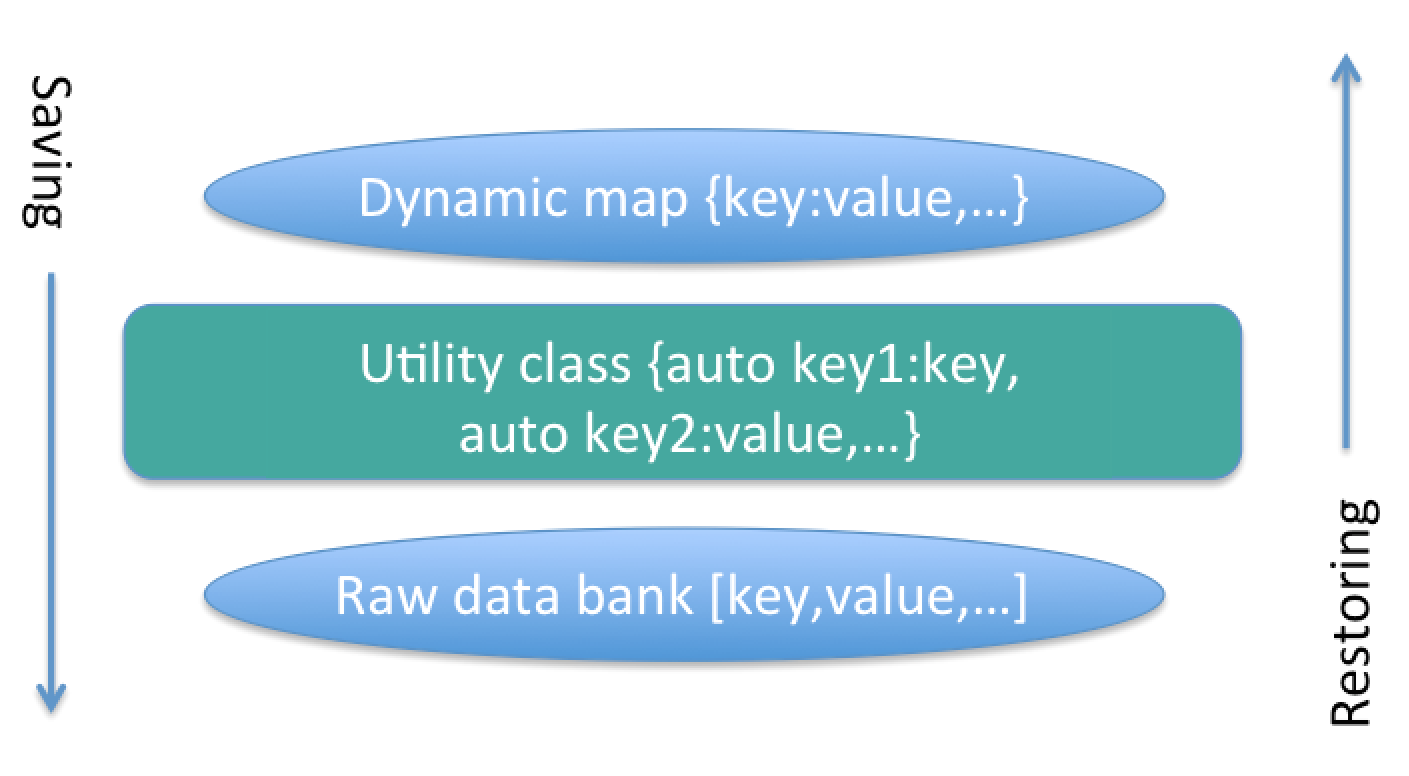}
\caption{Method used to save and restore dynamic maps.}
\label{fig:dynamic}
\end{center}
\end{figure}

While physics measurements mostly rely on information about the signal candidate,
 the complete event is often exploited to calculate quantities that 
discriminate against background processes.
A number of such quantities are employed in analyses, and the tools that 
calculate these quantities typically have parameters that are tailored to the needs of individual measurements.  
Examples of such quantities are isolation variables, which are measures of particle multiplicity within a region of the detector 
around triggered candidates. The necessary size of the region is analysis dependent, and some measurements use a set of regions with different sizes.
This is a different use case, in which a static map is no longer
appropriate to save and restore such data. In order to deal with such cases, an automatically
generated utility class key is created to store separately the map key and corresponding value to be saved.
This ensures that the key and data of the original dynamic map
are both written to the raw data stream, as shown in Figure~\ref{fig:dynamic}.

Extra objects needed for analysis, such as primary vertices that have been fitted without
the signal tracks, are persisted one level down from the top of the decay chain.
Since these objects represent static analysis level classes, the usual persistence approach is used
rather than the dynamic map.

\subsection{Analysis using the online reconstruction}

\begin{figure}[t]
\begin{center}
\includegraphics[width=0.45\textwidth]{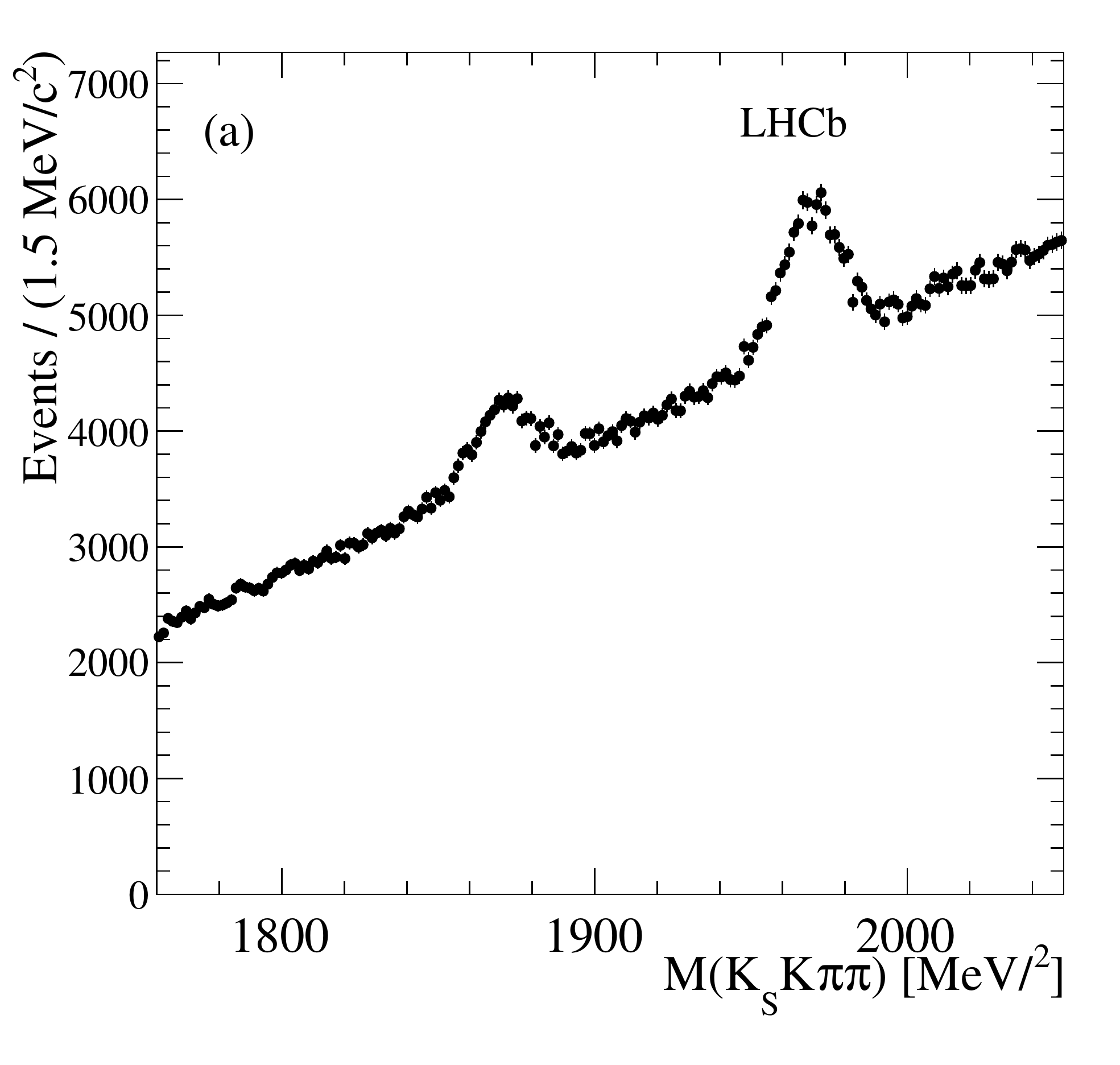}
\includegraphics[width=0.45\textwidth]{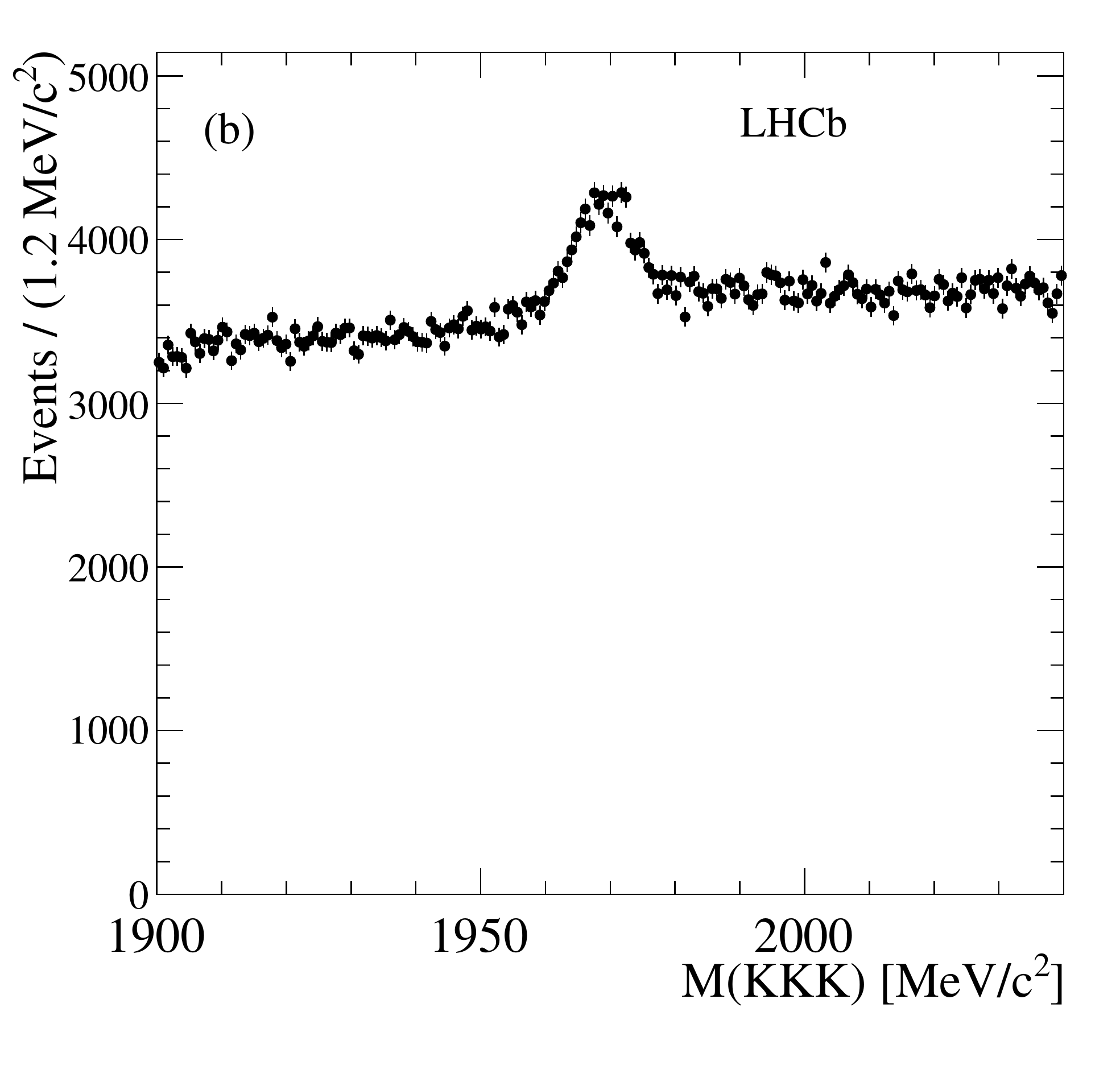}
\includegraphics[width=0.45\textwidth]{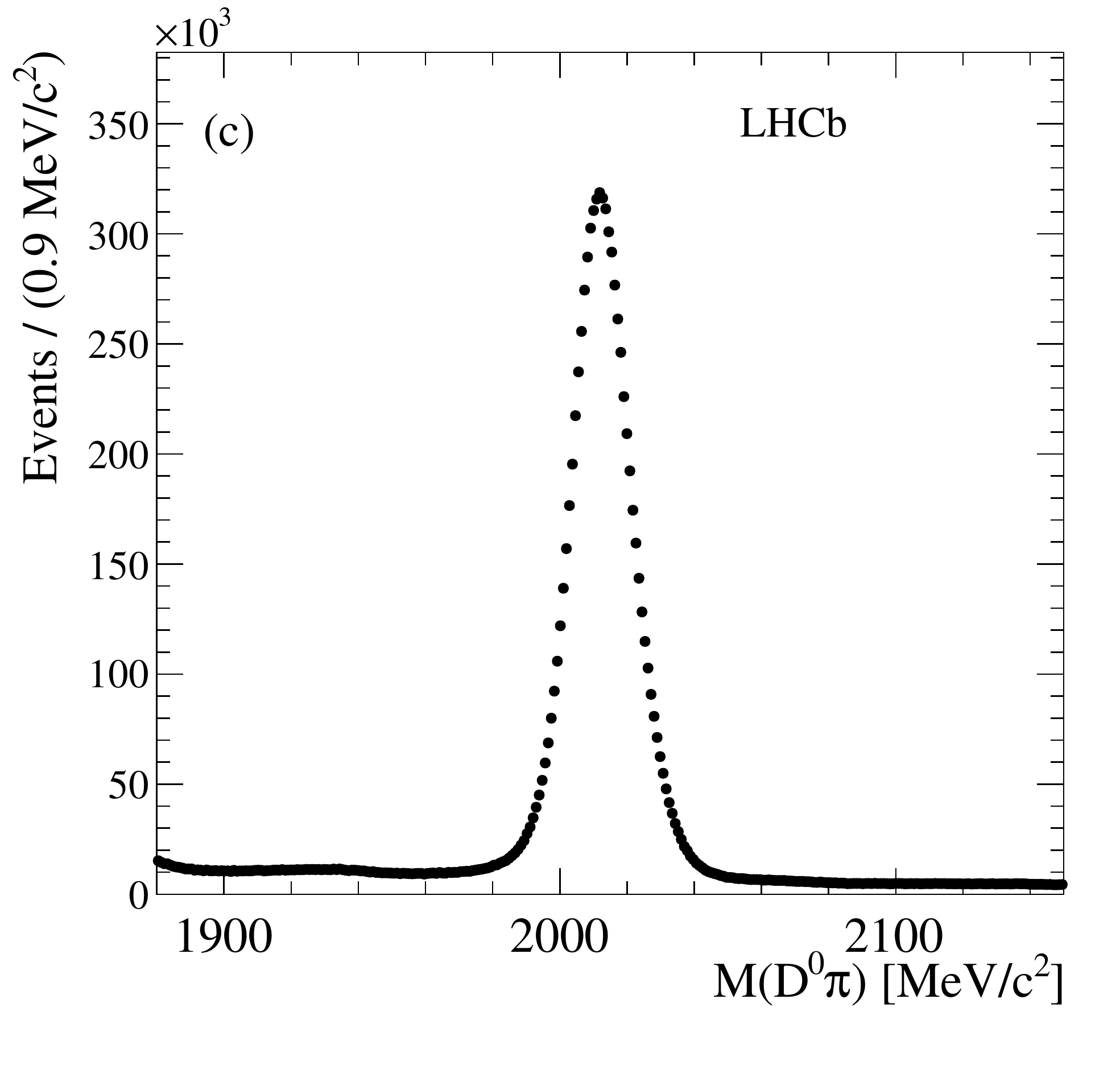}
\includegraphics[width=0.45\textwidth]{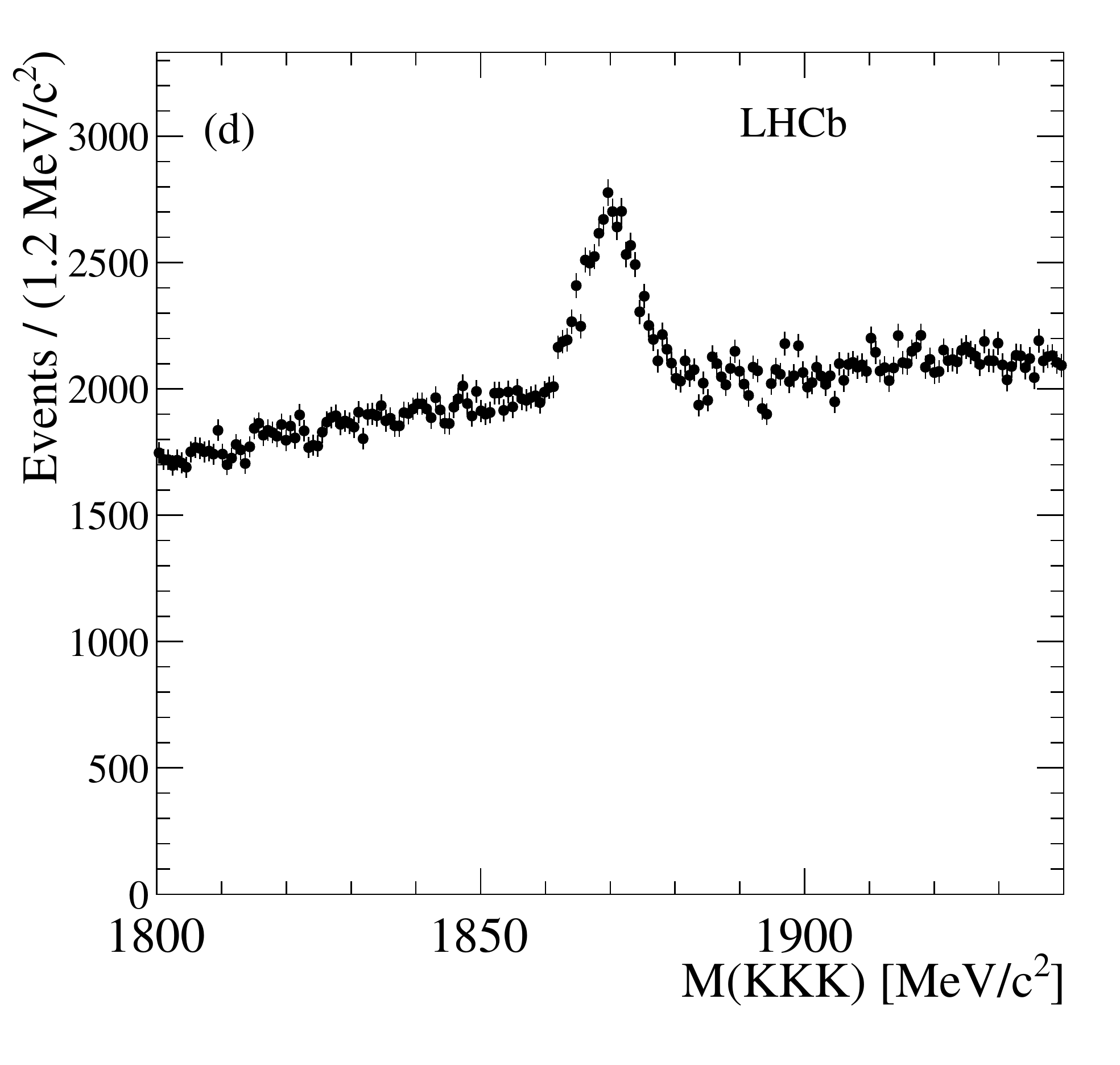}
\caption{
Invariant mass distributions for candidates reconstructed in the $\KS\Km\pip\pip$ (a), $\Km\Kp\Kp$ (b),
$\Dz(\to\Km\pip)\pip$ (c), and $\Km\Kp\Kp$ (d) state hypotheses, from 26.5\invpb of proton collision data taken in 
2015. The peaks corresponding to \Dp, \Ds, and \Dstarp mesons can clearly be seen.
}
\label{fig:lineshapes}
\end{center}
\end{figure}

The intended purpose of the Tesla application and the Turbo stream infrastructure is to ensure that 
more decays can be collected and therefore more precise measurements performed
than would have been possible under the Run-I computing model. 
Example data distributions showing clean decay signals directly from the information
of the online reconstruction are provided in Figure~\ref{fig:lineshapes}.
The limits on the output rate using the Full stream mean that only a subset
of the high-rate particle decays would have been collected without the Turbo
stream infrastructure.

\section{Outlook and future prospects}
\label{sec:future}

The use of the Turbo stream in 2015 proved to be successful. 
The first two published physics measurements from the LHCb experiment based 
on data collected in the 2015 run were based on the Turbo stream~\cite{LHCB-PAPER-2015-037,LHCB-PAPER-2015-041}. 
Around half of the HLT2 trigger lines currently persist
the trigger reconstruction  using the Turbo stream.

\subsection{Use in the upgraded LHCb experiment}

The upgrade of the LHCb experiment will see the luminosity rise from
$4\times10^{32}{\rm cm^{-2}s^{-1}}$ to $2\times10^{33}{\rm cm^{-2}s^{-1}}$.
In addition, the L0 hardware trigger will be removed. Therefore the software trigger
will need to cope with input rates that are \mysim30 times larger than the current trigger
input rate.
The output rate of $b$-hadrons, $c$-hadrons and light
long-lived hadrons will increase significantly, as shown in Table~\ref{tab:rate}.
\begin{table}[h]
\centering
\begin{tabular}{l|rr}
Particle type & Run I (kHz) & Upgrade (kHz) \\ \hline
$b$-hadrons & 17.3&270 \\
$c$-hadrons & 66.9&800 \\
light long-lived hadrons & 22.8& 264 \\
\end{tabular}
\caption{Rates of hadrons in Run I compared to the expected rate
in the upgraded LHCb experiment~\cite{anatomy}.}
\label{tab:rate}
\end{table}
In order to accommodate the increase in rate, the use of specific selection triggers
will become increasingly necessary. 
The Turbo model will become increasingly utilised as an increased retention will have a direct
impact on physics reach.
With the rate of interesting events being so high, the output bandwidth becomes a concern.
The average size of a raw event in the LHCb upgrade is anticipated to be $\mysim100$\,kB.
An event rate of 20\,kHz in traditional model will
then be 2\,GB/s. Assuming $1.5\times10^7$\,s of data taking each year, storage space at the level of
30\,PB/year would be required to save the data in raw form. Usually multiple copies of the raw data are stored to
ensure safety against data loss. In addition, the size of the reconstructed event is larger than the raw form, meaning the total
storage needs would be at the level of 300\,PB/year. 
The traditional model therefore does not scale to the LHCb upgrade regime.
The Turbo model would only require 100\,MB/s and would provide $20\times$ the output event rate 
of the traditional approach with the same resources, assuming all events
in the upgraded LHCb experiment use the Turbo model. 
A mixture of Turbo and traditional approaches is expected to be used as many
analyses require the use of full event information.

\subsection{Extended use of the trigger reconstruction}

\begin{figure}[t]
\begin{center}
\includegraphics[width=0.6\textwidth]{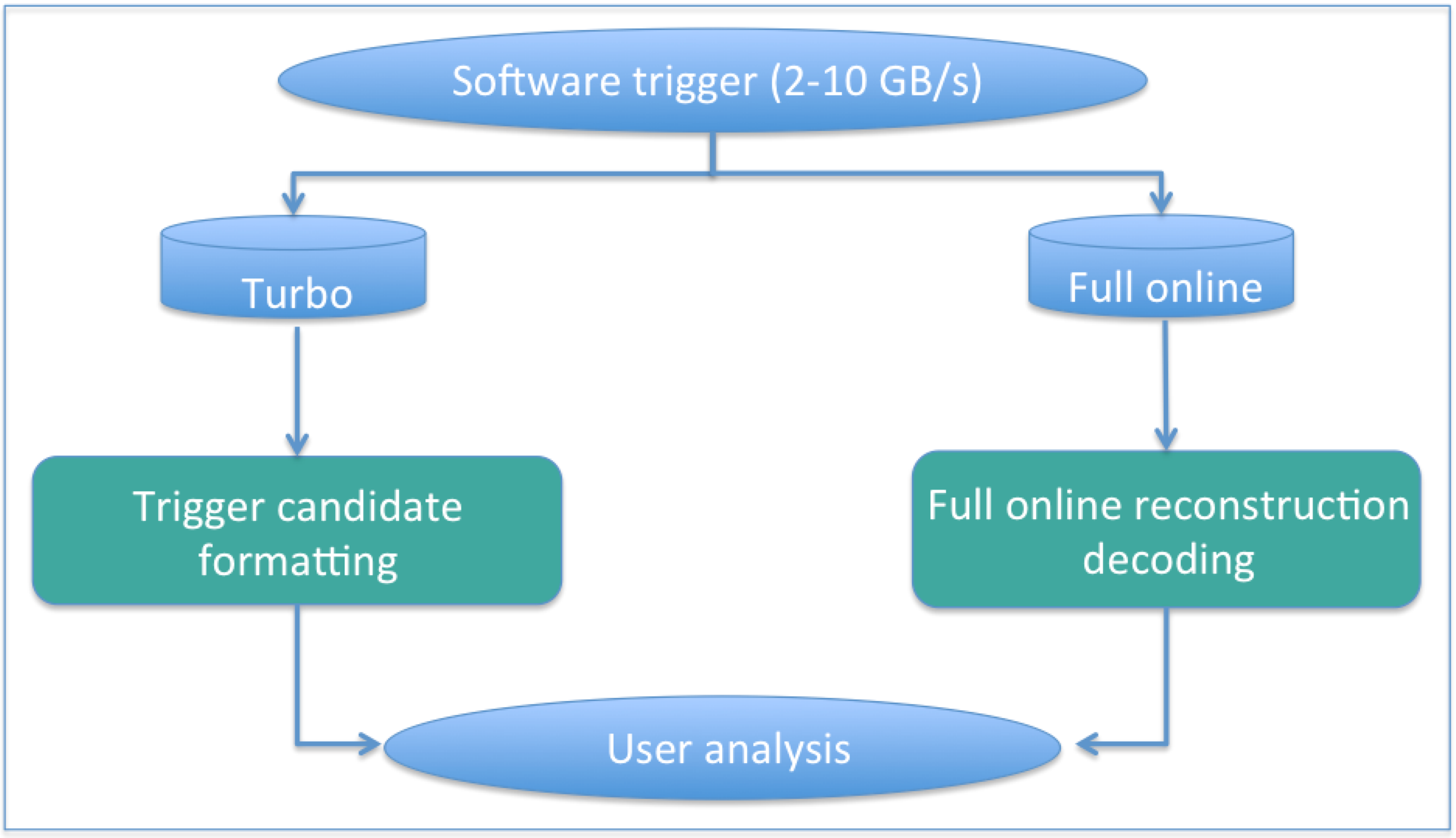}
\caption{Possible LHCb Upgrade data processing model.}
\label{fig:patternUp}
\end{center}
\end{figure}

The methods described in Sections~\ref{sec:run2} and~\ref{sec:implementation}
 use the trigger reconstruction to reduce the event size
thus relaxing the upper limit on the event rate.
The single reconstruction and the removal of a layer of selections also
simplify analyses, potentially reducing many sources of systematic uncertainty.
These benefits would also be seen if the complete reconstructed event in the trigger
was available for analysis. The event model is depicted in Figure~\ref{fig:patternUp}.
In a similar fashion to the Turbo model, where analysis classes for candidates are saved
inside the raw data banks, the physics objects created by the online reconstruction
may also be saved to the raw data bank. With a formatting level to prepare the data 
for user level analysis tools, an additional reconstruction can be avoided therefore
reducing the computing time.
With appropriate filters inside the trigger,
the decision of whether to keep the raw event or not could be made on an event-by-event
basis.

\section{Summary}
\label{sec:summary}

The Tesla toolkit allows for analyses to be based 
on the event reconstruction that is performed by the LHCb HLT.
By design, the Tesla output files are compatible with existing analysis framework software
with minimal changes required from analysts.

The event reconstruction performed by the HLT is of sufficient quality
for use in physics analyses because the detector is aligned and calibrated
in real time during data taking.
This is in turn made possible through the upgraded computing infrastructure introduced
in the first long shutdown of the LHC and the decision to buffer data from the first software trigger level.

The successful commissioning of this concept in 2015 has allowed multiple analyses to be performed
based on this model. The higher output rates of the LHCb upgrade will make the approach
increasingly necessary.

\section*{Acknowledgements}

\noindent 
We thank the technical and administrative staff at the LHCb
institutes. We acknowledge support from CERN and from the national
agencies: CAPES, CNPq, FAPERJ and FINEP (Brazil); 
CNRS/IN2P3 (France); BMBF, DFG and MPG (Germany); INFN (Italy); 
FOM and NWO (The Netherlands);  
SNSF and SER (Switzerland); 
STFC (United Kingdom).
We acknowledge the computing resources that are provided by CERN, IN2P3 (France), 
KIT and DESY (Germany), INFN (Italy), SURF (The Netherlands), PIC (Spain), GridPP (United Kingdom), 
RRCKI and Yandex LLC (Russia), CSCS (Switzerland), IFIN-HH (Romania), 
CBPF (Brazil), PL-GRID (Poland) and OSC (USA). We are indebted to the communities behind the multiple open 
source software packages on which we depend.

\addcontentsline{toc}{section}{References}
\ifx\mcitethebibliography\mciteundefinedmacro
\PackageError{LHCb.bst}{mciteplus.sty has not been loaded}
{This bibstyle requires the use of the mciteplus package.}\fi
\providecommand{\href}[2]{#2}

\newpage
\end{document}